\documentclass[12pt]{article} 
\usepackage{epsfig}

\newcommand{\go}{G\={o}}

\begin{document}
%\draft
\title{Unbiased simulation of \\structural transitions in calmodulin}
%\title{Simplified protein dynamics simulations exhibiting large-scale conformational transitions}
\author{Daniel M. Zuckerman \\
%}
%\address{
\normalsize
Department of Environmental \& Occupational Health, \\
\normalsize
Graduate School of Public Health, University of Pittsburgh, \\
\normalsize
and \\
\normalsize
Center for Computational Biology \& Bioinformatics, \\
\normalsize
University of Pittsburgh, 
Pittsburgh, PA 15213,  \\
\normalsize
dzuckerman@ceoh.pitt.edu
}
\date{\today}
\date{Draft: \today}
\maketitle

%\vspace{-1.5cm}
\begin{figure}[here]
\hspace*{-0.3in}
%\vspace*{2cm}
\epsfig{file=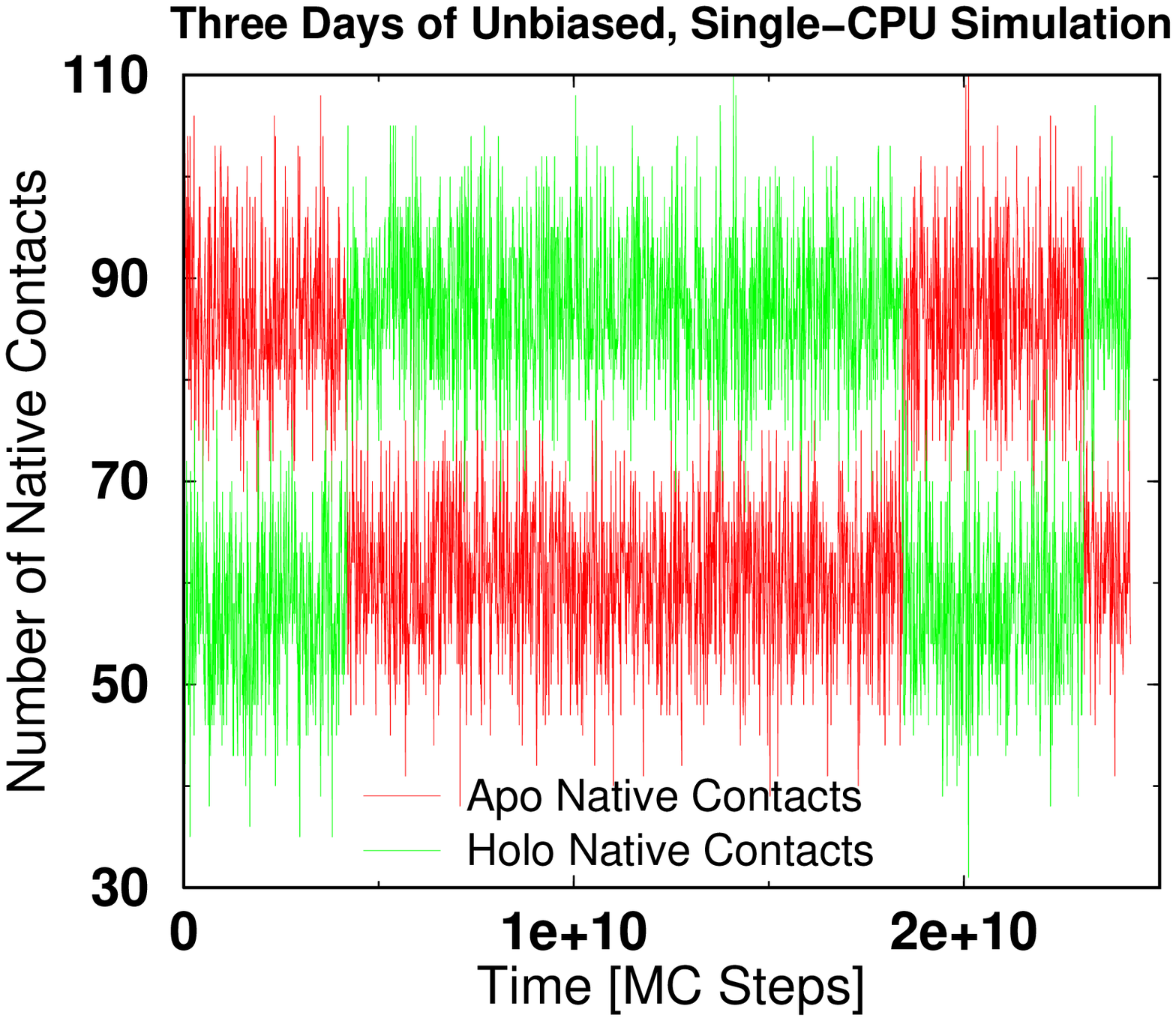, height=5in}
%\hspace*{-1in}
%\epsfig{file=Cam72intermediates-slide.eps, height=4.5in}
\end{figure}

\newpage
\begin{abstract}
We introduce an approach for performing ``very long'' computer simulations of the dynamics of simplified, folded proteins.
Using an alpha-carbon protein model and a fine grid to mimic continuum computations at increased speed, we perform unbiased simulations which exhibit many large-scale conformational transitions at low cost.
In the case of the 72-residue N-terminal domain of calmodulin, the approach yields structural transitions between the calcium-free and calcium-bound structures at a rate of roughly one per day on a single Intel processor.
Stable intermediates can be clearly characterized.
The model employs \go-like interactions to stabilize two (or more) experimentally-determined structures.
The approach is trivially parallelizable and readily generalizes to more complex potentials at minimal cost.
\end{abstract}

\section{Introduction}
The biological functions of many proteins result from their \emph{folded-state} dynamics, and in particular, from conformational changes among meta-stable states.
Motor proteins perform their vital functions based upon structural transitions, and additionally, such transitions often accompany ligand-binding and catalysis events:
those occurring in calmodulin and in adenylate kinase upon ligand binding are textbook examples \cite{Berg-2002} among numerous others. 
However, little is known about the detailed dynamics of these large-scale transitions.
How cooperative are the changes?
Do meta-stable intermediate states often occur?
How many reaction pathways play an important role?
Do any generic kinetic or dynamical features appear, especially in light of the expectation that nature tunes barrier heights to a few times the thermal energy scale $k_B T \sim RT$?

Calmodulin (CaM), a 148-residue calcium-binding and signalling protein, is an ideal test system because of its modest size and because it has been the subject of intensive experimental scrutiny (e.g., \cite{Cook-1988,Bax-1992,Quiocho-1992,Quiocho-1993,Bax-1995b,Tsien-1997,Akke-1999,Shea-2002,Vogel-2003}).
CaM undergoes large scale conformational transitions both within and between its two domains.
CaM has also been studied computationally, for instance by the Garc\'{i}a and Kuczera groups \cite{Garcia-2001,Kuczera-2002a,Kuczera-2002b}, but these all-atom molecular dynamics studies have been limited to nsec timescales.

Many approaches have been developed for the general problem of determining reaction pathways --- including both ``static,'' ``quasi-dynamic,'' and ``ensemble'' approaches. 
We term ``static'' those approaches which yield a single, presumably optimal pathway (e.g., \cite{Muller-1979,Berkowitz-1983,Wodak-1985,Elber-1989,Elber-1990,Elber-1991,Gillilan-1992,Karplus-1992,Theodorou-1993,Chirikjian-2002a,Chirikjian-2002b}).
The so-called static picture, by definition, does not account for inherent thermal fluctuations or the possibility of multiple pathways, though Elber and Shalloway modelled the effect of temperature in a static picture \cite{Elber-2000}.
Biased molecular dynamics approaches, like targeted and steered dynamics \cite{Harvey-1993,Wollmer-1993,Diaz-1997,Schulten-1997} attempt to include more realistic aspects of the true dynamics;
such methods typically generate a small number of potentially dominant dynamical pathways in protein systems, rather than a full ensemble.

``Ensemble'' approaches attempt to generate suitably distributed sets of reacting trajectories.
The pioneering work of Pratt employing the Metropolis approach \cite{Pratt-1986} was applied by Chandler and coworkers to a variety of problems (e.g., \cite{Chandler-1998a,Chandler-1998b,Chandler-1998c,Chandler-1998d,Chandler-1998e,Chandler-1998f}).
Elber and coworkers developed the stochastic difference equation approach, which generates approximate reactive trajectories without high-frequency motions \cite{Elber-1996,Elber-1999a,Elber-2002}.
Woolf and Zuckerman purused a non-Metropolis ensemble approach \cite{Woolf-1998e,Zuckerman-1999,Zuckerman-2001,soft-preprint}, as did Mazonka \emph{et al.} \cite{Mazonka-1998};
see also the work of Eastman \emph{et al.} \cite{Eastman-2000}.
A notable, independent approach for studying rare events is the ``weighted-ensemble Brownian dynamics'' of Huber and Kim \cite{Huber-1996}, which is conceptually related to the earlier work of Harvey and Gabb \cite{Harvey-1993}.
%Yet despite the variety and power of the biased ensemble methods, their predictions for larger-scale systems have never been rigorously checked due to the lack of prohibitively expensive unbiased trajectories.

\begin{figure}
\begin{center}
\epsfig{file=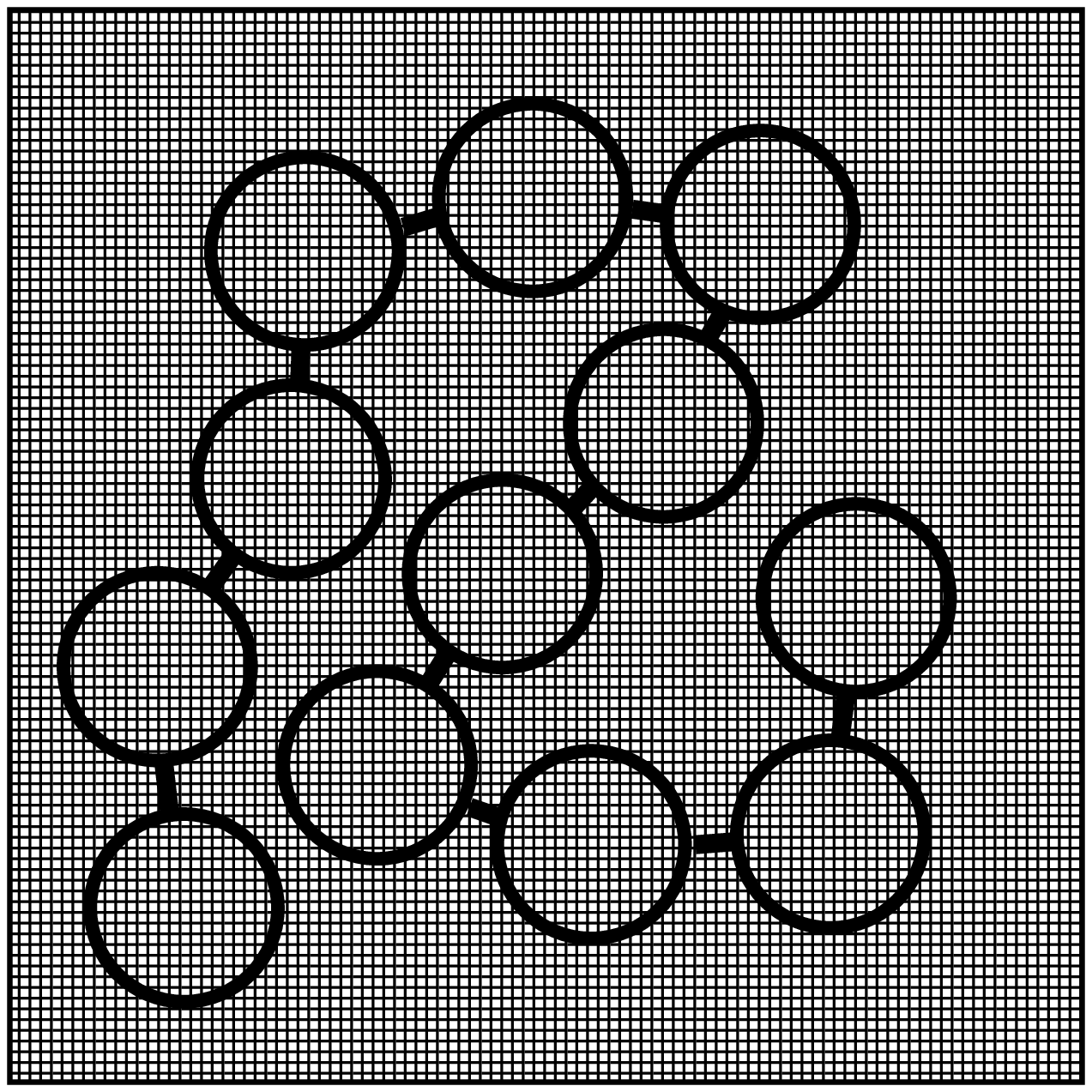, height=2.5in}
\epsfig{file=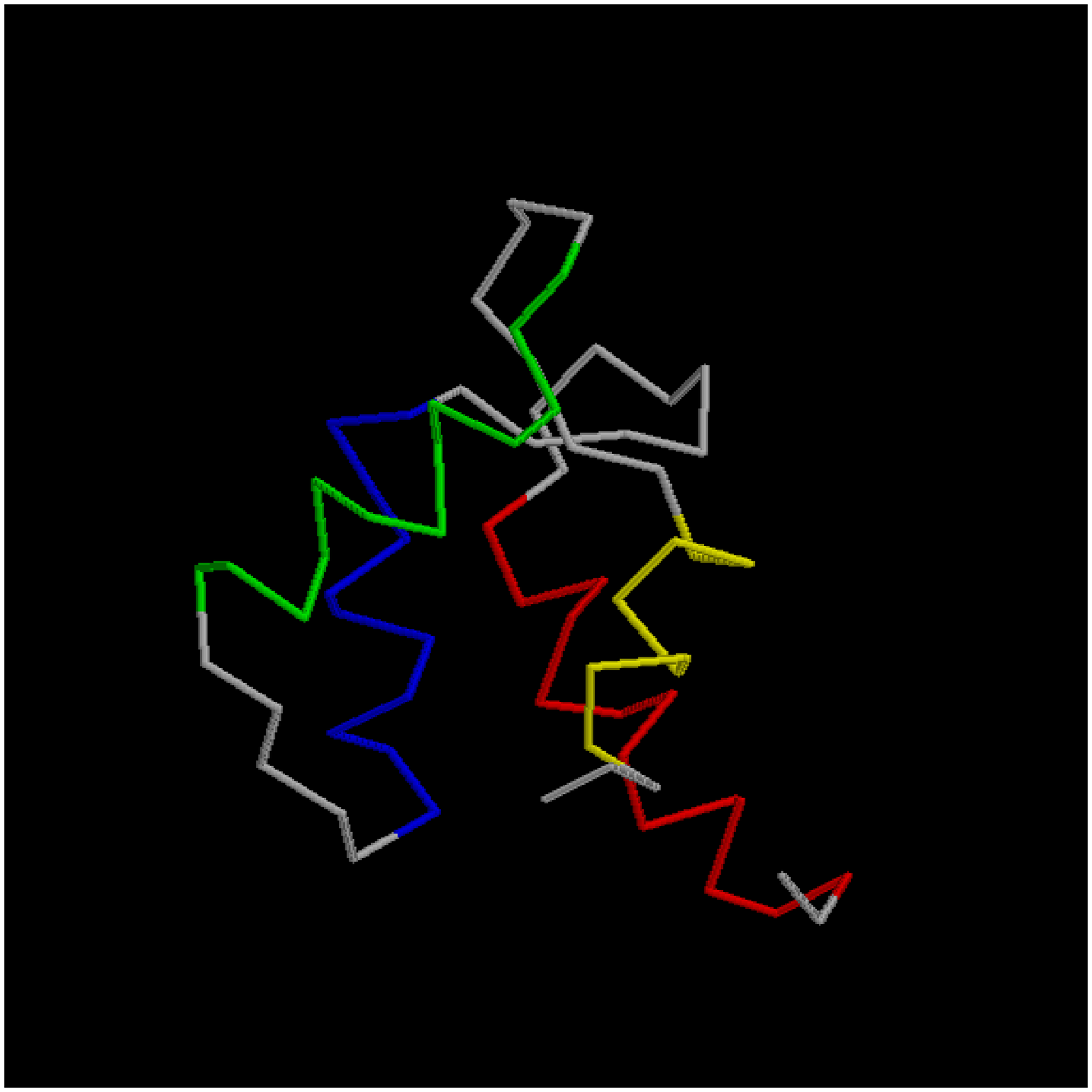, height=2.5in}
\caption{\label{fig:beads-on-grid}
A schematic bead model of a protein on a fine grid, and the residue-level depiction of calmodulin's N-terminal domain.
The centers of the residue beads (alpha-carbon locations) are restricted to grid sites.
Because the grid spacing is much less than the size of the beads, the discretized simulation mimics a continuum calculation. 
Interaction potentials for all possible bead orientations are stored in memory during the simulation, saving potentially costly calculation.
The right panel depicts an averaged simulation structure, starting from the configuration of PDB code 1cfd.
}
\end{center}
\end{figure}

One recent study pursued similar goals to the present investigation for a cubic-lattice toy model.
Specifically, Borovinskiy and Grosberg studied the design of cubic-lattice ``protein'' models capable of undergoing conformational transitions \cite{Borovinskiy-2002}.

Here we introduce an unbiased methodology to study large-scale structural transitions in folded proteins, and we use it to perform an initial examination of the structural dynamics in the N-terminal domain of calmodulin.
The method builds on two existing computational strategies.
The first, due to Panagiotopoulos and Kumar \cite{Panagiotopoulos-1999,Panagiotopoulos-2000}, is a general discretization approach which involves the use of a fine grid (with lattice spacing much less than the particle or atom size) to mimic continuum calculations.
Atoms or particles are only permitted to occupy grid sites, and interaction energies or forces are computed in advance for all necessary inter-particle displacements; see Fig.\ \ref{fig:beads-on-grid}.
Panagiotopoulos and Kumar found that critical and coexistence properties of fluids could be reproduced very precisely with a speed gain of one to two orders of magnitude \cite{Panagiotopoulos-1999,Panagiotopoulos-2000}.
The notion of using a fine grid to simulate biomolecules also conceptually draws from coarse-grained lattice simulations of proteins \cite{Go-1978,Skolnick-1988,Skolnick-1994,Skolnick-1996,Dill-1989a,Dill-1989b,Dill-1990,Jernigan-1990}, particulularly the high coordination lattice studies by Kolinski, Skolnick, and coworkers.

The second principal computational precursor to the present work is the \go\ model \cite{Go-1975,Go-1978}, which represents proteins in solely a structural way, without reference to the underlying chemistry.
In particular, the \go\ model employs a chain of residue ``beads'', each of which only reacts favorably with other residues which are nearby in the \emph{native} structure.
Non-native contacts are penalized.
This simple model was developed to study the dynamics of protein folding and it is still widely used for that end \cite{Clementi-2000,Clementi-2000b,Shakhnovich-2002}.
Note that the \go\ model is trivially generalized to other levels of chemical detail --- e.g., to all-atom simulation \cite{Shakhnovich-2002}.

While the present approach is not restricted to simplified models, we use a highly reduced (residue-level) model both because it is easier to implement and because residue-level models have already provided meaningful bio-macromolecular results.
Reduced models of proteins have often been used in the past, particularly to study protein folding dynamics and thermodynamics (e.g., \cite{Levitt-1975,Levitt-1976,Miyazawa-1985,Miyazawa-1996,Skolnick-1988,Skolnick-1994,Skolnick-1996,Doniach-1989,Wolynes-1989,Wolynes-1991,Thirumalai-1990,Thirumalai-1992,Thirumalai-1992b,Karplus-1996,Karplus-1997,Karplus-1999,Clementi-2000,Clementi-2000b,Hall-2001a,Hall-2001b,Banavar-2001}), ``ab initio'' protein strucutre prediction (e.g., \cite{Scheraga-1976,Scheraga-1997a,Scheraga-1997b,Scheraga-2001,Friesner-1994,Friesner-1994b,Friesner-1995,Elber-2000b,Rose-1995}), and coarse-grained dynamics (e.g., \cite{Bahar-1997c});
see also a review of structure-based potentials \cite{Jernigan-1996}.
Moreover, it has been shown conclusively that even models which do not distinguish among atom or amino acid types can capture intermediate scale (i.e., alpha carbon fluctuations) and large scale motions (i.e., the slowest modes) \cite{Tirion-1996,Bahar-1997a,Bahar-1997b,Bahar-1998,Bahar-1999,Hinsen-1998}.

The present, unbiased approach should also prove useful in the refinement of the biased ``ensemble'' methods discusssed above.
Because the ensemble methods have not been fully vetted in protein systems, the present approach can usefully contribute ``perfect'' ensembles of \emph{unbiased} large-molecule transition trajectories which can serve as ``gold standards'' for comparison with biased methods.
In particular, unbiased ensembles can help shift the focus to a number of biologically and methodologically important questions regarding intermediates:
(i) can their lifetimes be estimated accurately?
(ii) do intermediates introduce a very wide variation in the durations of transition events? 
(iii) are multiple pathways typically observed?

In outline, the next section, \ref{sec:method}, describes the simulation approach used here.
We emphasize that our method is not tied to the choice of model (i.e., forcefield), which is highly simplified for this initial study and is described in Sec.\ \ref{sec:model}.
Sec.\ \ref{sec:results} presents the simulation results, and highlights the straightforward observation of stable intermediates.
Our concluding discussion is given in Sec.\ \ref{sec:conclude}.

\begin{figure}[t]
\begin{center}
\epsfig{file=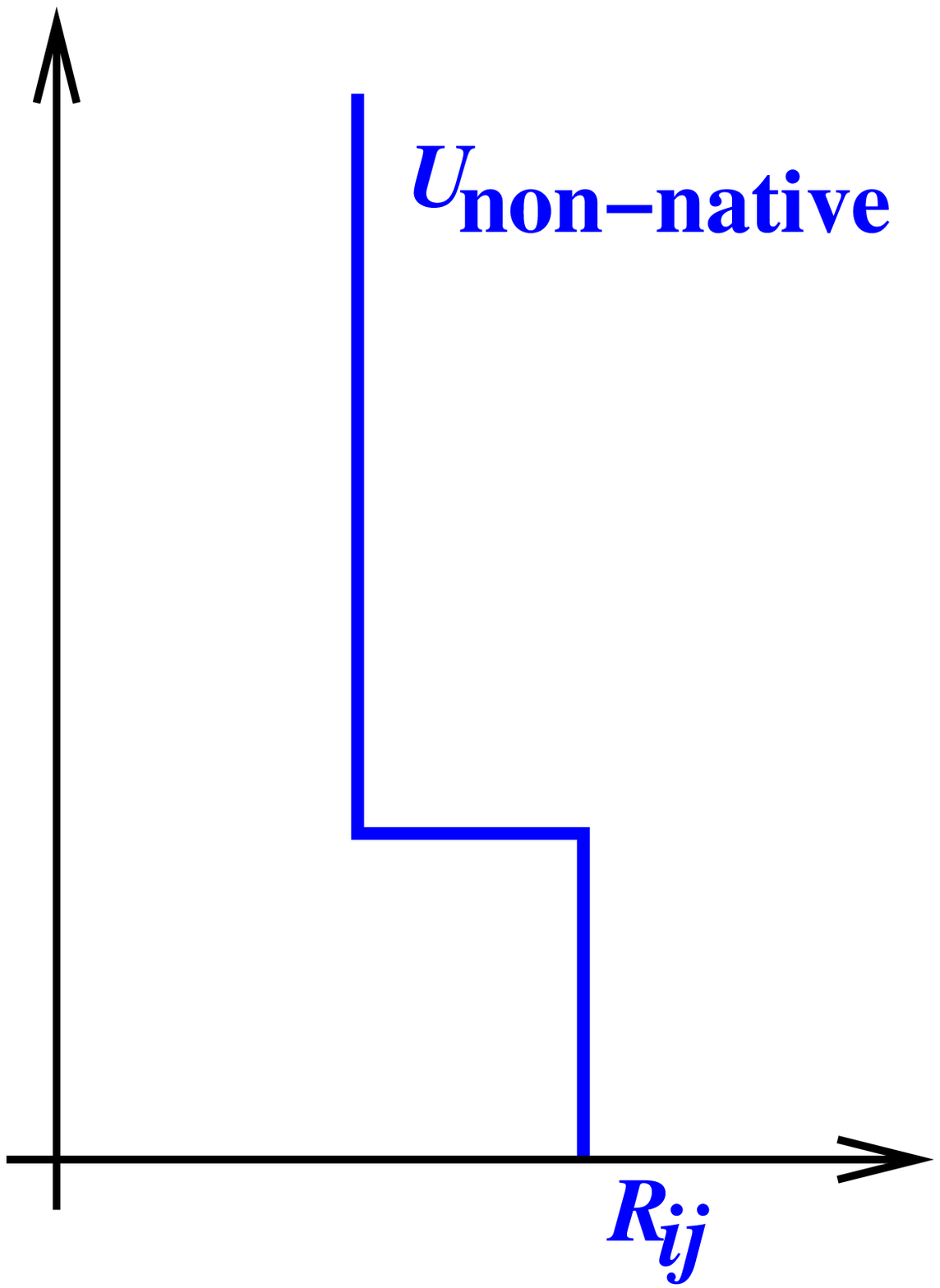, height=2in}
\epsfig{file=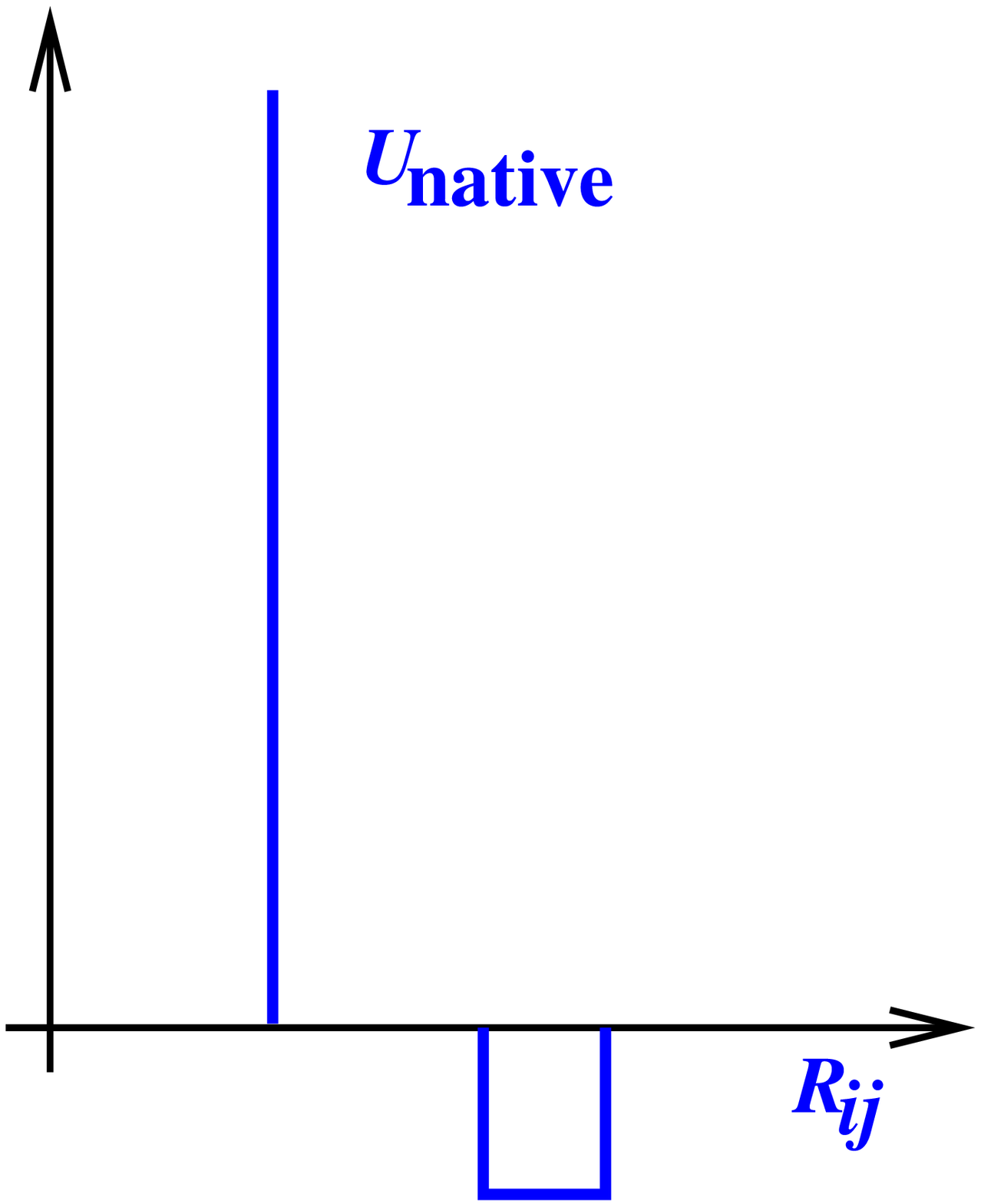, height=2in}
\epsfig{file=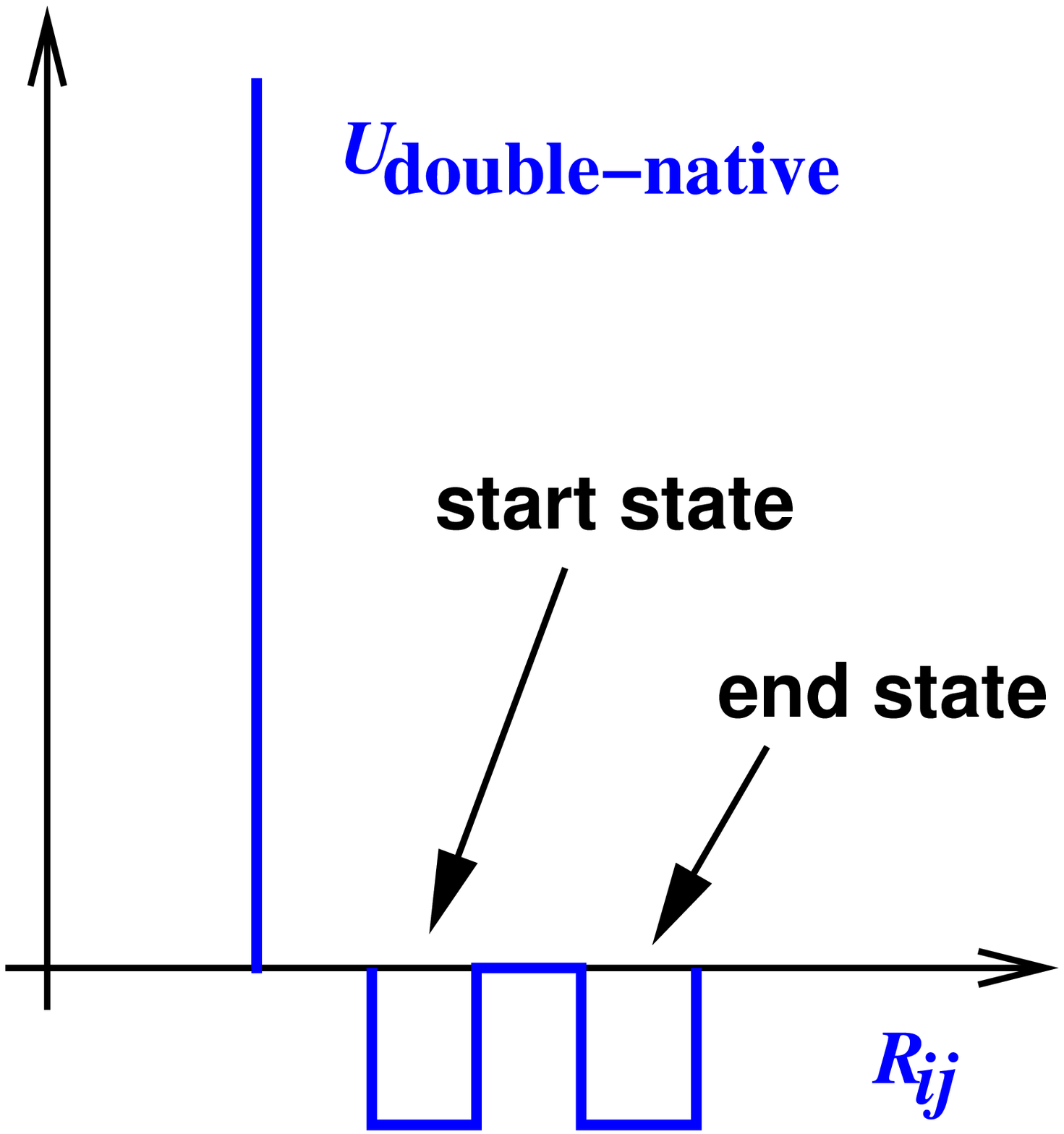, height=2in}
\caption{\label{fig:dbl-go-potl}
The highly reduced, \go-like, inter-residue potentials used in the simulations.
The ``non-native'' pair potential is used between residues that are not in contact in the \emph{holo} or the \emph{apo} structures of calmodulin.
A contact is defined by a distance $R_{ij} < 8$ \AA\ between alpha-carbon atoms of residues $i$ and $j$ in the Protein Data Bank coordinate files.
The ``single-native'' potential applies for residue pairs in contact in only one of the two structures, while ``double-native'' is for pairs contacting in both.
}
\end{center}
\end{figure}

\section{Methodology}
\label{sec:method}

\subsection{Fine-grid simulation}
The fine-grid simulation methodology is adopted directly from Panagiotopoulos and Kumar \cite{Panagiotopoulos-1999,Panagiotopoulos-2000}.
The idea is to allow particles (here, residues) to occupy only discrete positions on a grid.
Interaction energies are then stored in arrays, rather than computed at each dynamics step.

More specifically, a ``fine'' grid is used --- i.e., one in which the grid spacing is much smaller than the particle size --- in an attempt to mimic continuum results.
This point bears emphasizing: 
the grid approach is adopted solely as a means to reproduce continuum results at greater computational speed;
the errors or artifacts resulting from using a fine grid are expected to be negligible due to the small size of the grid spacing.
Indeed, the central result of Panagiotopoulos and Kumar \cite{Panagiotopoulos-1999,Panagiotopoulos-2000} is that very sensitive liquid-vapor coexistence curves and critical points could be determined quite accurately within the grid approach.
A rough rule of thumb from Panagiotopoulos' work is that a lattice spacing of $1/5$ the particle size is sufficient for practical computations.

How fine a grid spacing is necesary in the present context?
Owing to the crudeness of the models used here (see Sec.\ \ref{sec:model}), we do not require high precision.
However, to implement a dynamic Monte Carlo scheme (see below) with a sufficient acceptance ratio, a small grid spacing is required. 
In the simulations discussed here, the lattice spacing was 0.13 \AA, which may be compared with residue-residue interaction distances of 4 - 8 \AA. 
In other words, a very fine grid is employed here.

\begin{figure}[t]
\begin{center}
\epsfig{file=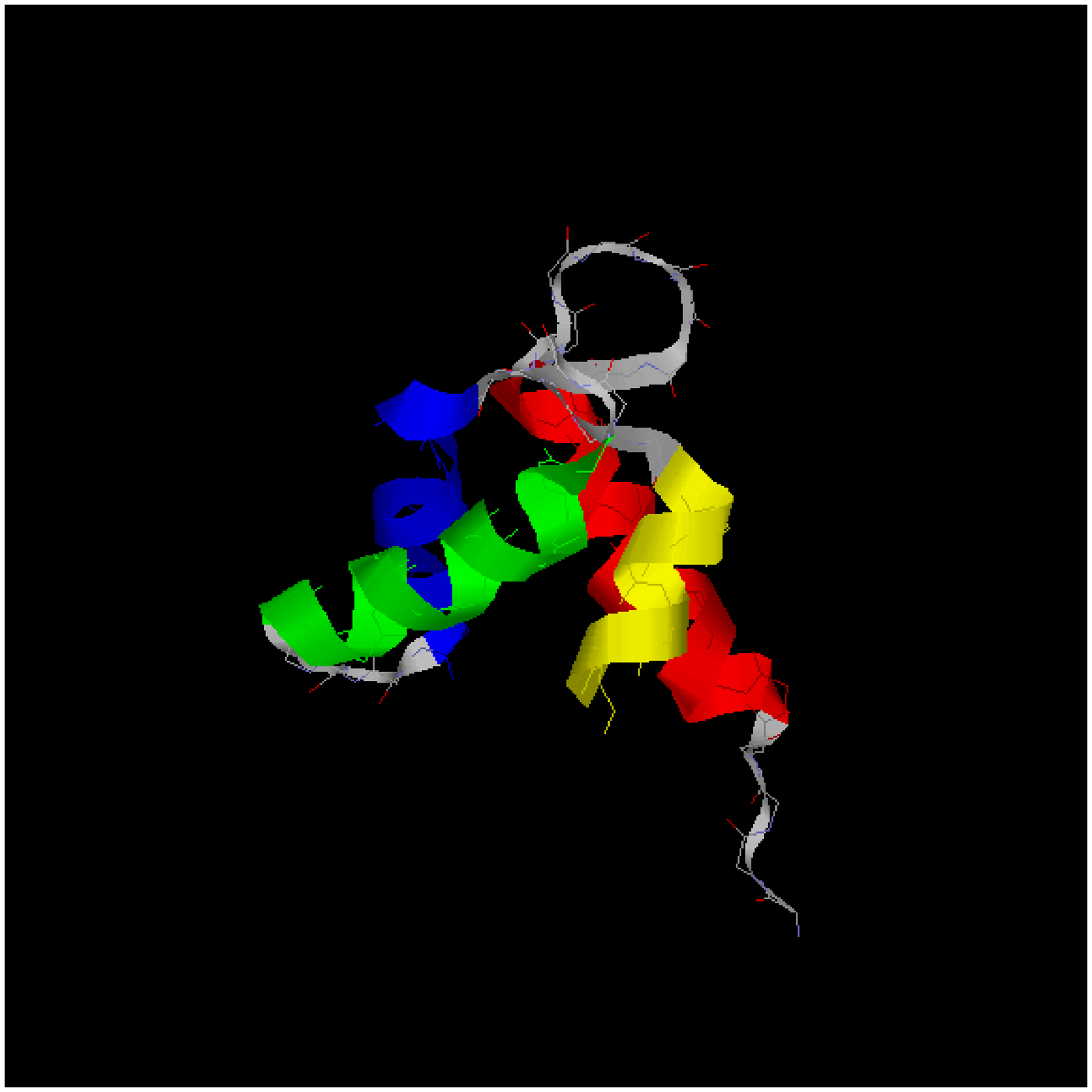, height=2.5in}
\epsfig{file=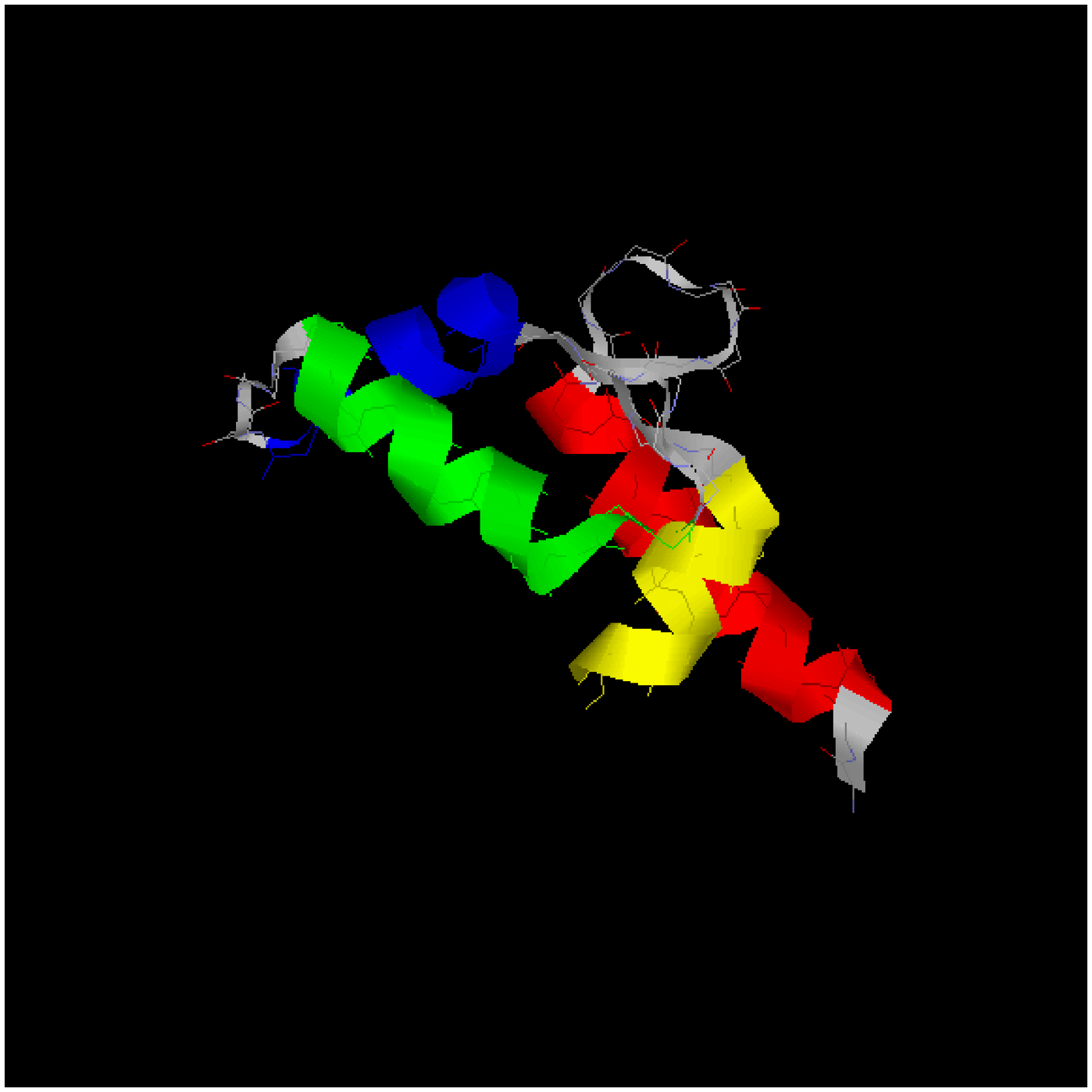, height=2.5in}
\caption{\label{fig:cam-structures}
The ``start'' and ``end'' calmodulin structures used to construct the \go-like potential.
The left panel is the calcium-free (\emph{apo}) structure, PDB code 1cfd, determined by solution NMR, from which the simulations were started.
The right panel depicts the calcium bound (\emph{holo}) strucutre, PDB code 1cll, determined from X-ray diffraction.
}
\end{center}
\end{figure}

\subsection{Dynamic Metropolis Monte Carlo}
In this initial implementation, we have chosen to use dynamic Monte Carlo (DMC) dynamics, which is a common, current choice for simulations of protein folding dynamics (e.g., \cite{Clementi-2000,Clementi-2000b,Shakhnovich-2002}).
While DMC dynamics were chosen because of the relative ease of implementation, they may be justified on two grounds.
(i) Because only small, local trial moves will be considered --- see below --- and accepted with a Boltzmann-factor-preserving probability, DMC may be considered a variant of overdamped Brownian dynamics.  
Brownian dynamics have a well-understood physical basis \cite{Allen-Tildesley}.
(ii) Further, due to the highly simplified nature of the models used here (see Sec.\ \ref{sec:model}), one can expect the simulated dynamics --- of any kind --- only to give a qualitative picture.
This qualitative description should not be sacrificed by Brownian-like dynamics embodied in DMC.

The dynamic Monte Carlo is designed to make highly local moves in a trivial way: 
at every time step, a randomly chosen residue makes a trial move with uniform probability to one of the 26 lattice sites on the surface of the $3 \times 3 \times 3$ cube centered at its present location.
The move is accepted or rejected according to the usual Metropolis criterion: accept if and only if the energy decreases or a random number $R_u$ chosen uniformly from the interval $0<R_u\leq1$ is less than $\exp{( -\Delta U / RT )}$, where $\Delta U$ is the change in the total energy (Hamilitonian) of the system. 

\subsection{Hardware, Software, and Computer Time}
The results reported below are based on a computer program written in the C language and compiled with the ``gcc'' compiler on Linux machines.
Simulations were run on Intel processors of both 2.4 and 2.8 GHz, which permitted approximately 3 - 4 $\times 10^8$ Monte Carlo steps per hour.
All the trajectories discussed here were obtained in less than one week of of processor time.

\section{Model}
\label{sec:model}

The residue-level model describes a highly reduced protein without any true chemistry, following the spirit of many previous workers \cite{Go-1975,Go-1978,Tirion-1996,Bahar-1997a,Hinsen-1998}.
Nevertheless, the model does include the essential toplogical features of calmodulin: 
residue connectivity and sterics, as well as the attractive interactions found in both \emph{apo} and \emph{holo} strucrures.
Only the N-terminal domain was included --- specifically the 72 residues numbered 4 - 75 in PDB structures 1cfd (\emph{apo}, unbound) and 1cll (\emph{holo}, bound to calcium ions).

The model of the N-terminal domain of calmodulin is specified by covalent, steric, and attractive interactions;
see Fig.\ \ref{fig:dbl-go-potl}.
Consectutive residues are ``covalently'' connected by an infinitely deep, attractive square well centered on the native-state separation distance of structure 1cfd;
the well width was taken to be 10\% of the native separation ($\pm$5\%). 
Sterically, each residue excludes other residues from its hard-core exclusion zone, a sphere with radius set to 47.5\% (i.e., 95\%/2) of the minimum distance to other non-consecutive residues in the 1cfd native state.
Finally, \go-like attractive interactions are represented by square wells (of uniform depth $\epsilon$) centered at residue separation distances ($\pm$5\%) of less than 8 \AA\ from \emph{either} of the native structures (i.e., PDB codes 1cfd or 1cll).
All simulations were performed at a temperature given by 
$k_B T = 0.6 \epsilon$.

The double-well, attractive interactions depicted in Fig.\ \ref{fig:dbl-go-potl} merit further justification.
At first glance, such a potential may seem unnatural, particularly to those readers familiar with the distinction between the pair potential energy and potential of mean force (see, e.g., \cite{Mcquarrie}):
for a simple fluid consisting of spherical particles, it is well known that a single-well potential gives rise to a multi-well potential of mean force.
However, amino acids are not spherical, and if one insists on a point-residue model (as here), the only way to represent direct interactions of differing orientations is for the potential to possess more than one well.
A second, perhaps facile, justification comes from the results:
the double well potential accomplishes the principal aim of stabilizing two distinct conformational states.
Indeed, it requires roughly a day of computer time to generate a fluctuation sufficiently large as to jump to the second state.
A third justification is more pragmatic:
when more realistic potentials (e.g., \cite{Miyazawa-1985,Miyazawa-1996,Karplus-1997,Karplus-1999,Scheraga-1997a,Scheraga-1997b,Clementi-2000,Clementi-2000b}) are discretized with the find-grid approach, presumably the ``unnatural'' double-\go\ interactions can be reduced to minimal levels required to stabilize the distinct states.

\begin{figure}
\begin{center}
\epsfig{file=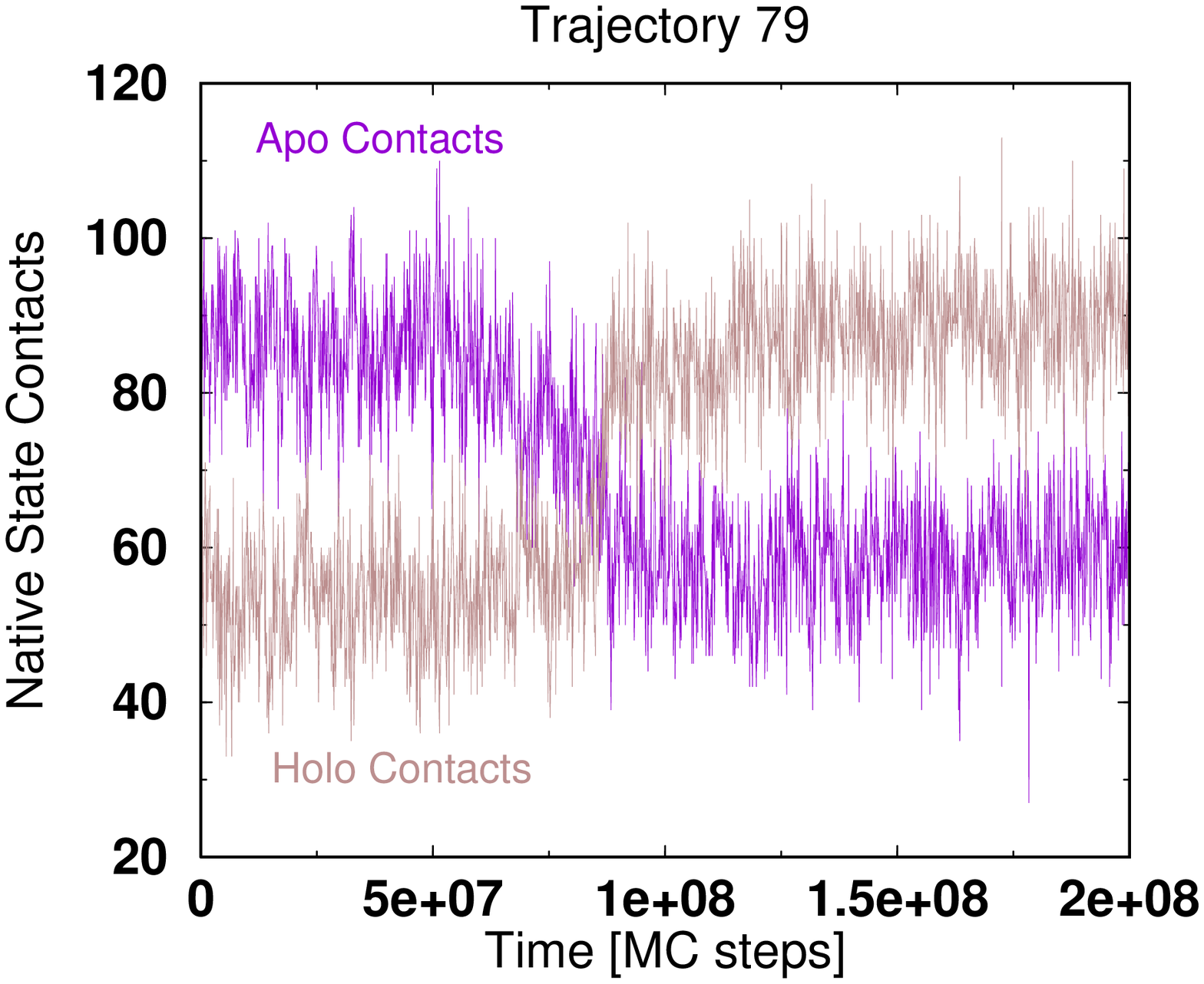, height=2.in}
\epsfig{file=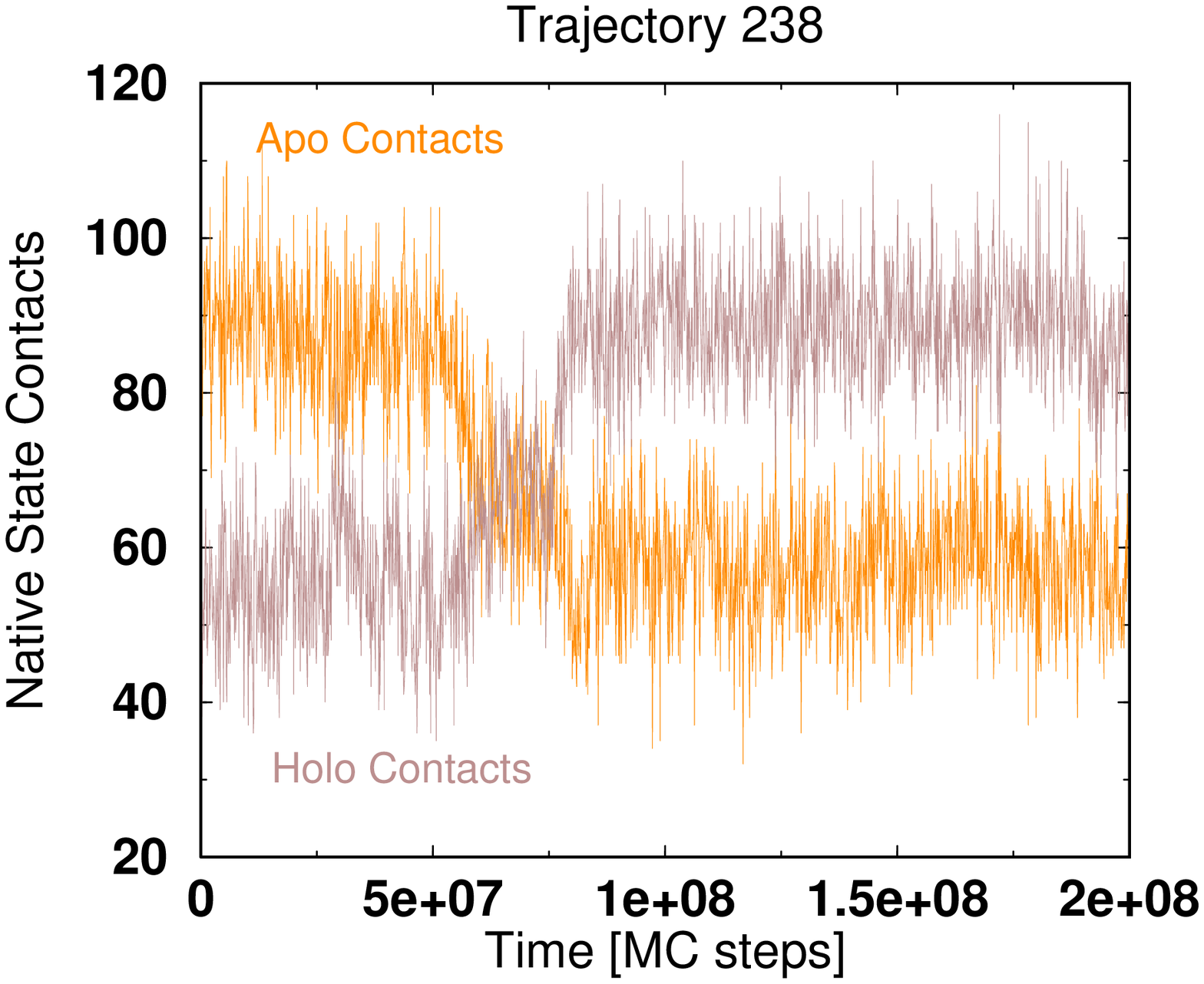, height=2.in}
\caption{\label{fig:sample-transns}
Two sample structural transitions in the simplified N-terminal domain of calmodulin, monitored by the numbers of native contacts in the \emph{apo} and \emph{holo} states.
Contacts are defined by inter-C$^\alpha$ distances of less than 8 \AA\, and native contacts are based on comparison to the PDB structures 1cfd (\emph{apo}) and 1cll (\emph{holo}).
Each of the two trajectories shown required roughly 30 minutes of single-processor (2.8 GHz) computer time, and such transition events occur roughly once per day on a single processor.
}
\end{center}
\end{figure}

\section{Results}
\label{sec:results}
\begin{figure}[here,b]
\begin{center}
\epsfig{file=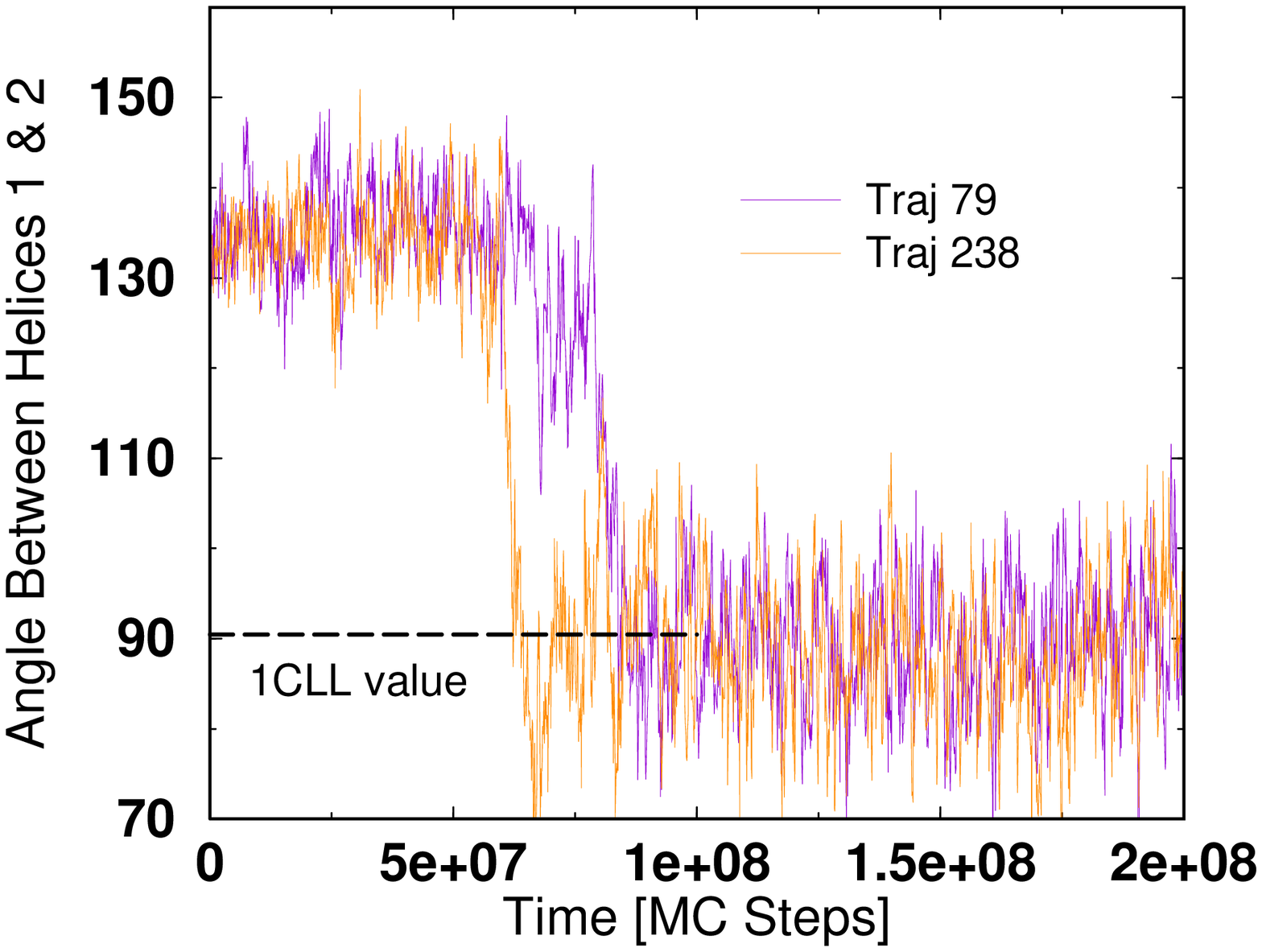, height=2.in}
\epsfig{file=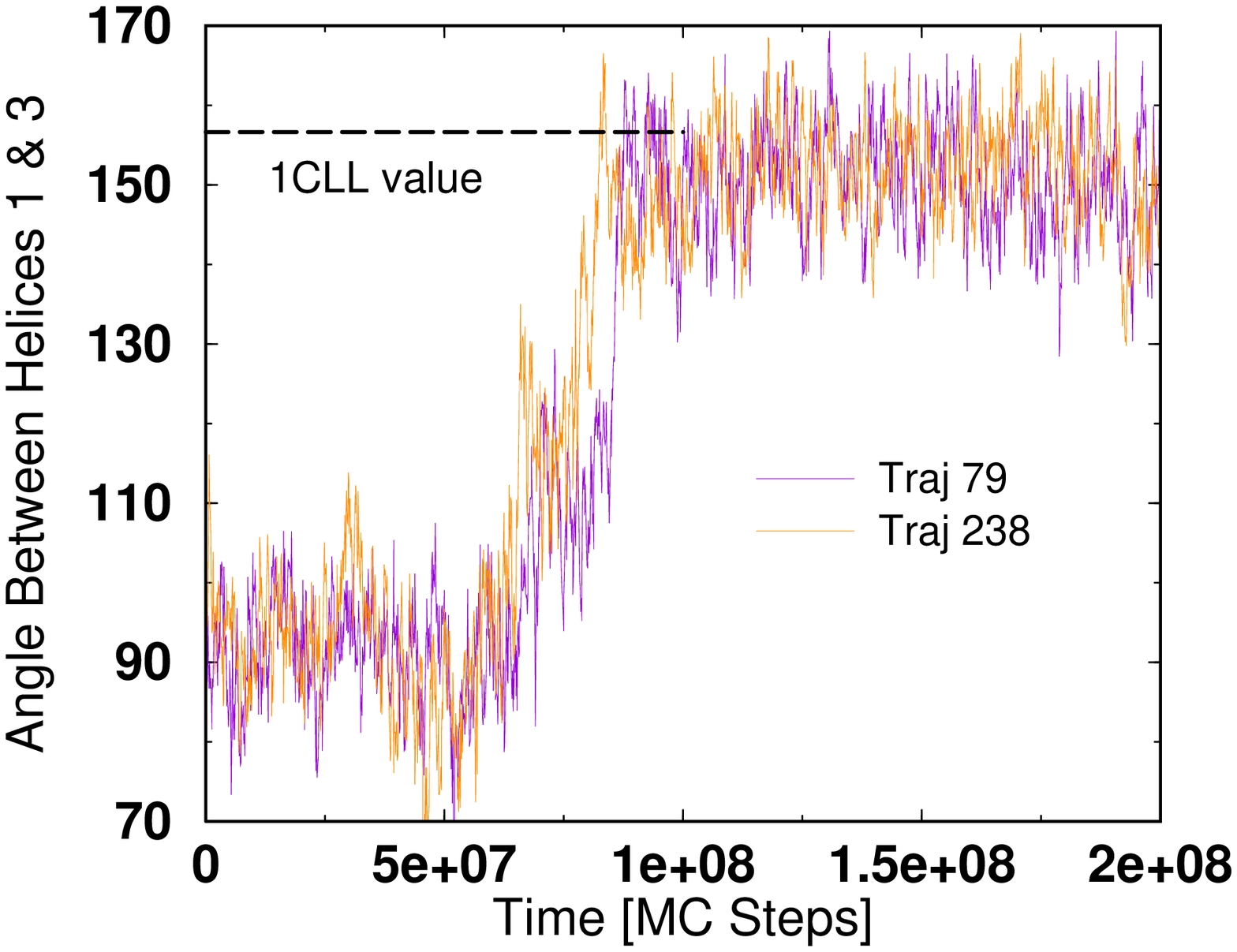, height=2.in}
\caption{\label{fig:sample-transns-angles}
The two transitions of Fig.\ \ref{fig:sample-transns} confirmed by comparison to the inter-helix angles in the PDB structure 1cll.
Helix axes were defined to be the vectors associated with the smallest moment of inertia of the residues present in the native structures.
}
\end{center}
\end{figure}

\begin{figure}[here]
\hspace*{-0.5in}
\epsfig{file=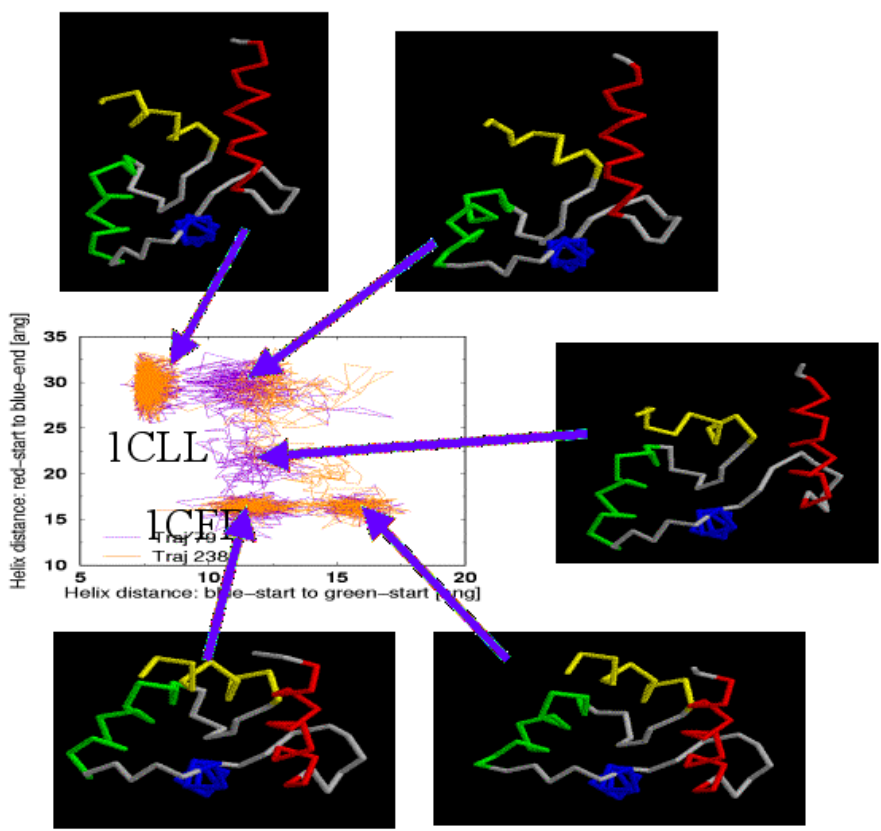, height=4.5in}
\caption{\label{fig:intermediates}
Long-lived intermediates in the transition events of Fig.\ \ref{fig:sample-transns}.
The central panel shows parametric traces of two independent transition trajectories, with the axes representing distances between the indicated helix ends;
note that the N-terminal ``start'' of the peptide is the red helix.
The surrounding structures are averaged over $10^7$ MC steps, but remained localized at the indicated positions.
Note that the two trajectories take slightly different pathways in the region where vertical axis ranges between 18 and 25;
the lower-right intermediate appears to be ``on-pathway'' for the orange trajectory but not for the purple.
}
\end{figure}

\begin{figure}
\begin{center}
\epsfig{file=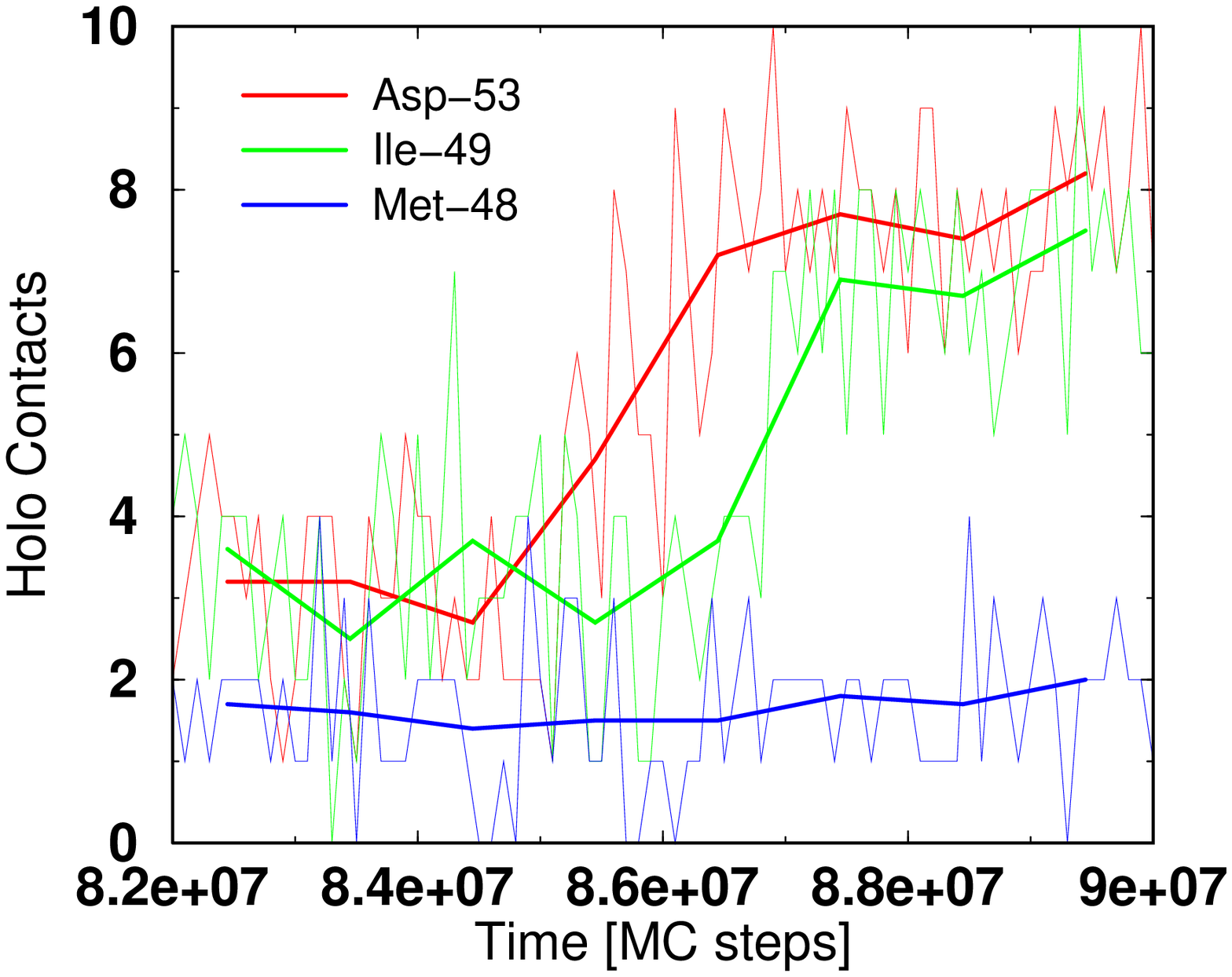, height=2.in}
\epsfig{file=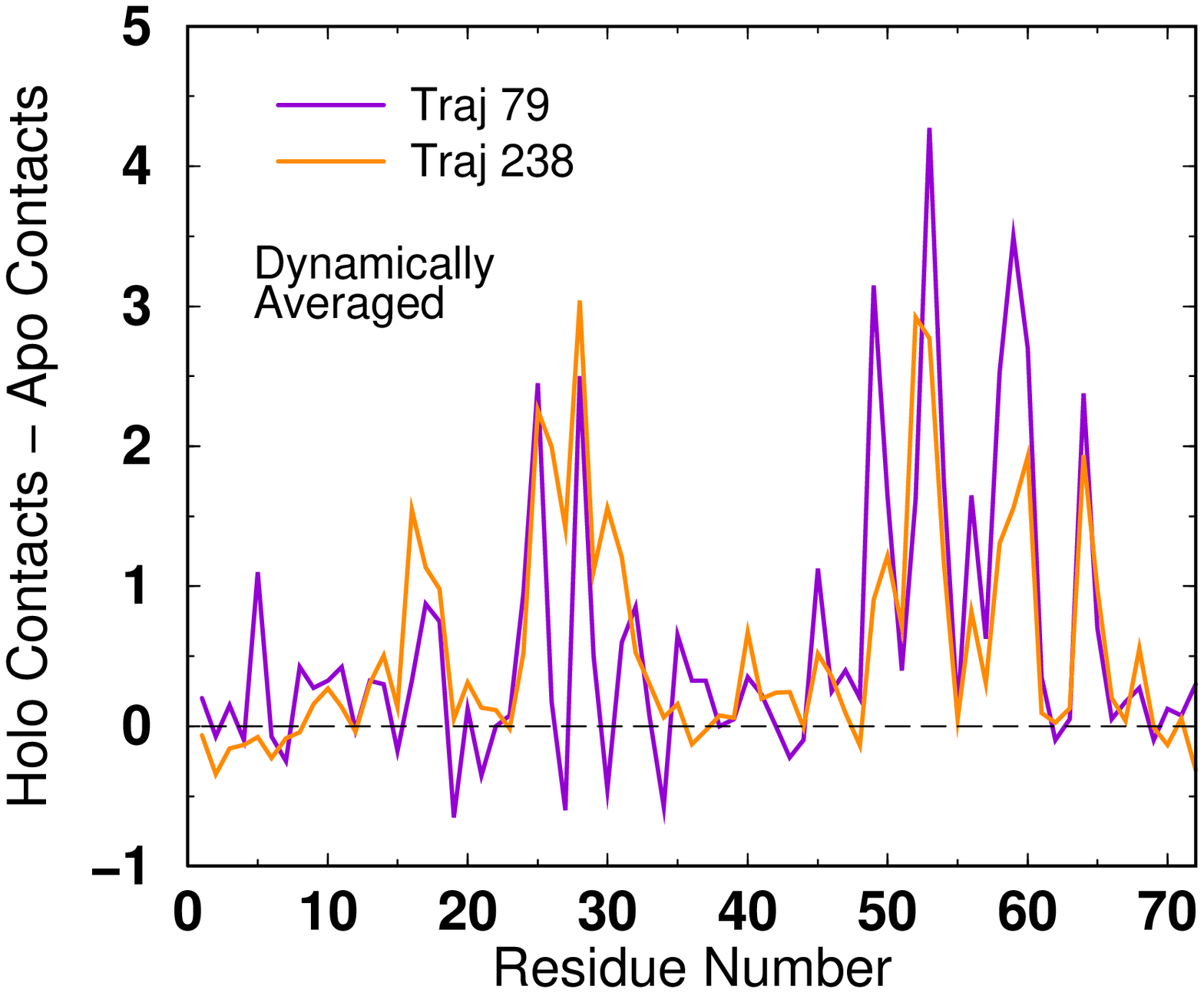, height=2.in}
\caption{\label{fig:res-contacts}
Dynamical behavior of residue contacts.
The left panel depicts the evolution of the contacts of three residues during the transition event of trajectory 79 (see Fig.\ \ref{fig:sample-transns}).
The ``\emph{holo} contacts'' for residue $i$ is the count of all residues which are both within $R_c = 8$ \AA\ of residue $i$ at the time point shown, and which are also within $R_c$ of $i$ in the \emph{holo} structure, 1cfd, depicted in Fig.\ \ref{fig:cam-structures}. 
Smoothed versions of the residue contacts evolution are also shown on the left for visual clarity.
The right panel shows the average differences in \emph{holo} contacts before and after the transition point at $8.6 10^7$ MC steps, for every residue in both transition of Fig.\ \ref{fig:sample-transns}.
Note the similarity of the traces for the two transitions, which --- interestingly --- \emph{are not correlated with contact-count differences based on the static PDB structures.}
}
\end{center}
\end{figure}

In this preliminary report, all the results are depicted in Figs.\ \ref{fig:sample-transns} - \ref{fig:res-contacts}, and described in the figure captions.
To summarize, sample transition events are shown (Fig.\ \ref{fig:sample-transns}), and traces of the helix angles confirm that the transition are genuine and sustained (Fig.\ \ref{fig:sample-transns-angles}).
Next, long-lived intermediates are readily visualized using compound variables which simply measure distances between helix ends (Fig.\ \ref{fig:intermediates}),
and finally the dynamical evolution of individual residue contacts is examined (Fig.\ \ref{fig:res-contacts}).

\section{Summary and Conclusions}
\label{sec:conclude}
We have introduced an approach for studying the long-time dynamics of large-scale conformational transitions in proteins. 
The technique was tested on a highly reduced, residue-level model of the N-terminus of calmodulin, which undergoes a dramatic rearrangement of its four helices  upon the (un)\-binding of calcium ions.
Such rearrangements were observed approximately once per day on a single processor in \emph{unbiased} dynamic Monte Carlo simulation.
The scheme is quite promising because it proved capable of spanning the range of timescales from that for inter-residue vibrations to the ``waiting time'' between transition events --- i.e., the inverse rate.

The approach is readily extendable in two important ways.
First, it is embarrassingly parallelizable, so an inexpensive, modest-sized Beowulf cluster can readily improve upon the single-processor output by an order of magnitude or more.
Second, the fine-grid discretization approach \cite{Panagiotopoulos-1999,Panagiotopoulos-2000} --- which simply mimics continuum calculations --- lends itself to more complex potentials:
while the addition of particles (e.g., ``beads'') to any model entails a cost, the use of a more chemically realistic potential will require minimal additional overhead.
Furthermore, other models --- reduced or atomistic --- can be stabilized in experimentally determined structural states using the same \go-like interactins employed here. 

Despite the apparent simplicity of the model employed here, the simulated structural transitions in calmodulin exhibited highly complex behavior.
Distinct transition pathways and quasi-stable intermediate states could be readily identified, as could critical residues.
The transition details identified in the present study cannot immediately be identified with expectations for experimental outcomes because of the simplicity of the model.
Nevertheless, the results should capture those aspects of the dynamics governed by sterics and connectivity, in analogy with the success of network-like models in predicting large-scale fluctuations \cite{Tirion-1996,Bahar-1997a,Bahar-1997b,Bahar-1998,Bahar-1999,Hinsen-1998}.

Future studies with more chemically realistic models (e.g., \cite{Miyazawa-1985,Miyazawa-1996,Rose-1995,Karplus-1997,Karplus-1999,Scheraga-1997a,Scheraga-1997b,Clementi-2000,Clementi-2000b,Hall-2001a,Hall-2001b}) implemented in the rapid, fine-grid approach should suggest testable experimental hypotheses. 
Note that the effects on dynamics of varying forcefields have recently been addressed by Freed and coworkers \cite{Freed-2003}.
A related question, which also can be addressed directly in the present scheme, regards the effects of \emph{dynamics type} --- e.g., Metropolis, Langevin, molecular dynamics --- on transition events. 

The future determination of reaction pathways and dynamics in a hierarchy of models will also address a question both scientifically and methodologicallly critical:
\emph{What is the minimal model necessary to capture residue-level features of structural transitions in proteins?}
From the ``scientific'' perspective, a minimal model will suggest which interactions govern the transition events,
while optimal methods should use that model which is least costly, computationally.

Finally, it is appropriate to ask what the unbiased transitions observed in the present study presage for the use of the biased ``ensemble'' methods discussed in the introduction.
The observation of multiple pathways and widely disparate transition event durations suggests that great care will be required in any biased effort to generate ensembles of transition trajectories. 
Nevertheless, the present approach should provide critical input for testing biased methods in protein systems --- namely, unbiased trajectories to be used for comparison.
From a practical standpoint, a set of reduced-model transition events may also provide useful starting points to sidestep the potential for trapping associated with the Pratt approach.

\section*{Acknowledgments}
The author has benefitted greatly from discussions with many scientists:
Ivet Bahar, David Deerfield, Jeffery Evanseck, William Furey, Hagai Meirovitch, Robert Swendsen, Dror Tobi.
Special thanks are due Professor Bahar for her invaluable advice, support, and encouragement.
%?Evanseck,Madura,Day

%\bibliographystyle{unsrt}
%\bibliography{dmz}

\end{document}